\input{aipcheck.tex}

  \def\selectedoptions{final}

\documentclass[
   \selectedoptions
  ]
  {aipproc}

\def\lsim{\mathrel{\lower2.5pt\vbox{\lineskip=0pt\baselineskip=0pt
           \hbox{$<$}\hbox{$\sim$}}}}
\def\gsim{\mathrel{\lower2.5pt\vbox{\lineskip=0pt\baselineskip=0pt
           \hbox{$>$}\hbox{$\sim$}}}}
\def\Lsun{\hbox{L$_{\odot}$}}

\layoutstyle{8x11single}

\SetInternalRegister\hbadness{8000} 

\newcommand\doingARLO[2][]{%
  \ifx\mmref\undefined #1\else #2\fi
}

\begin{document}

\title 
      [Constraints on the Accuracy of Photometric Redshifts 
Derived from BLAST and Herschel/SPIRE Sub-mm Surveys]
      {Constraints on the Accuracy of Photometric Redshifts 
Derived from BLAST and Herschel/SPIRE Sub-mm Surveys}


\author{Itziar Aretxaga}{
  address={INAOE, Aptdo. Postal 51 y 216, 72000 Puebla, Mexico},
  email={itziar@inaoep.mx}
}

\iftrue
\author{David. H. Hughes}{
  address={INAOE, Aptdo. Postal 51 y 216, 72000 Puebla, Mexico}
}

\author{Edward Chapin}{
  address={INAOE, Aptdo. Postal 51 y 216, 72000 Puebla, Mexico},
}

\author{Enrique Gazta\~naga}{
  address={INAOE, Aptdo. Postal 51 y 216, 72000 Puebla, Mexico},
}
\fi

\copyrightyear  {2001}

\begin{abstract}
More than 150 galaxies have been detected in blank-field millimetre
and sub-millimetre surveys. However the redshift distribution of
sub-mm galaxies remains uncertain due to the difficulty in identifying
their optical-IR counterparts, and subsequently obtaining their
spectroscopic emission-line redshifts. In this paper we discuss
results from a Monte-Carlo analysis of the accuracy with which one can
determine redshifts from photometric measurements at
sub-millimetre-FIR wavelengths. The analysis takes into account the
dispersion in colours introduced by including galaxies with a
distribution of SEDs, and by including photometric and absolute
calibration errors associated with real observations. We present
examples of the probability distribution of redshifts for individual
galaxies detected in the future BLAST and Herschel/SPIRE surveys. 
We show that the combination of BLAST and 850$\mu$m observations 
constrain the photometric redshifts with sufficient accuracy to 
pursue a program of spectroscopic follow-up with the 100m GBT.

\end{abstract}

\date{\today}

\maketitle

\section{DESCRIPTION OF THE TECHNIQUE AND RESULTS}


Determining the density of star formation as a function of redshift is
the primary science objective of the Balloon-borne Large Aperture
Submillimetre Telescope (BLAST, \cite{devlin}) and other sub-mm/mm
facilities.  Using Monte-Carlo simulations that take into account
realistic photometric and absolute calibration errors for future 
BLAST surveys, we show that it is possible to determine redshifts from
BLAST data (at 250, 350 and 500$\mu$m)
with a $1\sigma$ average precision of $\Delta z \sim \pm 0.6$.
A similar level of redshift accuracy is
found for simulated observations with the Herschel/SPIRE camera which will
operate at identical wavelengths to BLAST.

The power of this simple technique to derive redshifts arises from the
unique ability of BLAST and SPIRE observations to bracket the
ubiquitous rest-frame FIR peak (at $\sim 60-150\mu$m) in the spectral
energy distribution (SED) of high-redshift ($1
\le z \le 4$) galaxies undergoing a significant amount of star
formation.

To determine the accuracy of this method we have generated mock
catalogues of galaxies between $z=0$ and $z=6$ using an evolving
$60\mu$m luminosity function \cite{saunders} that reproduces the
observed 850$\mu$m number counts.  The sub-mm flux densities, and
colours, of these mock galaxies are calculated from SEDs selected at random from a library of template
starburst galaxies, ULIRGs and AGN (Fig.1). Observational noise is
then added to the intrinsic fluxes: 1$\sigma$ photometric errors of 5
and 2.5 mJy for the BLAST and Herschel/SPIRE observations
respectively, and in both cases an absolute calibration error of 7\%.

We are therefore able to determine the photometric redshift
probability distribution for any galaxy detected in BLAST surveys 
by comparing its {\em measured} BLAST colours 
with the complete distribution of {\em
simulated} colours and redshifts of galaxies in
the mock catalogue (see Fig.2).

It is instructive to illustrate the discrepancy between the
photometric redshifts determined from the colours 
of galaxies detected in our simulated BLAST (or SPIRE) surveys and
their true mock catalogue redshifts.  Figure 3 shows that over the
entire redshift range of our simulations, $0 \le z \le 6$, the
$1\sigma$ error in the photometric redshifts, $\Delta z$, derived from
detections in 3 BLAST filters is $\pm 0.6$.  However beyond $z \sim 4$
the BLAST data alone systematically underestimate the redshifts,
confusing $z \sim 4.5$ galaxies with $z \sim 2.5$  galaxies for
example, as the longest wavelength BLAST filter (500$\mu$m) moves
short-ward of the rest-frame FIR peak.

A significant improvement in the derived photometric redshifts
occurs when the BLAST (or SPIRE) observations are complemented with
data from a longer-wavelength survey ({\it e.g.}  ground-based
850$\mu$m SCUBA data).  This extension to the wavelength coverage
(250--850$\mu$m) ensures that the photometric redshifts are
uniformly distributed about the $z = z_{phot}$ regression line over
the entire range $0 < z < 6$, with an improved average error of $\Delta
z \sim \pm 0.4$ (Fig.~3). This method therefore continues to provide un-biased
estimates of photometric redshifts for the most distant galaxies.
Furthermore, an increased 
sensitivity in the observations (with SPIRE for instance)
naturally translates into an increased accuracy of the redshift
distributions.

%

\begin{figure}
  \includegraphics[height=0.5\textheight,angle=90]{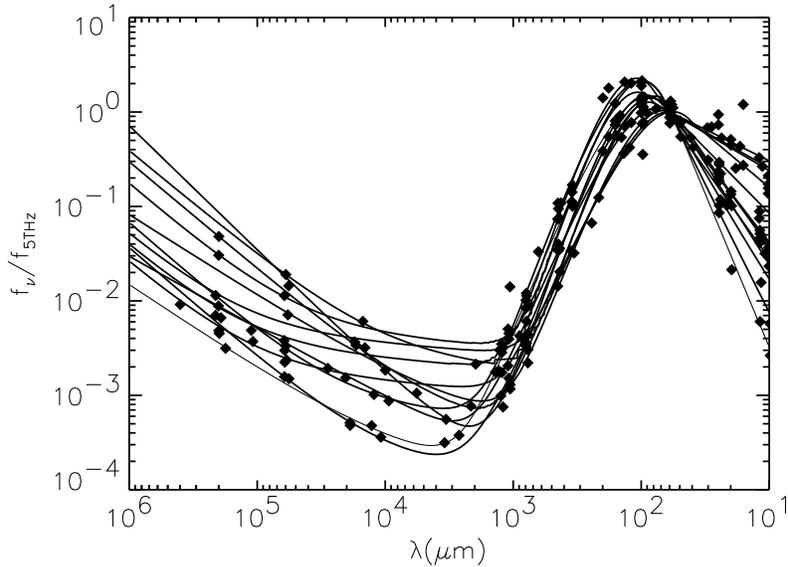}
  \caption{Rest-frame spectral energy distributions of 13 starburst
  galaxies, ULIRGs and AGN, normalized at 60$\mu$m.  Lines represent
  the best fit models to the SEDs and include contributions from
  non-thermal synchrotron emission, free-free and grey-body thermal
  emission.}
\end{figure}

\begin{figure}
  \includegraphics[height=0.4\textheight,angle=90]{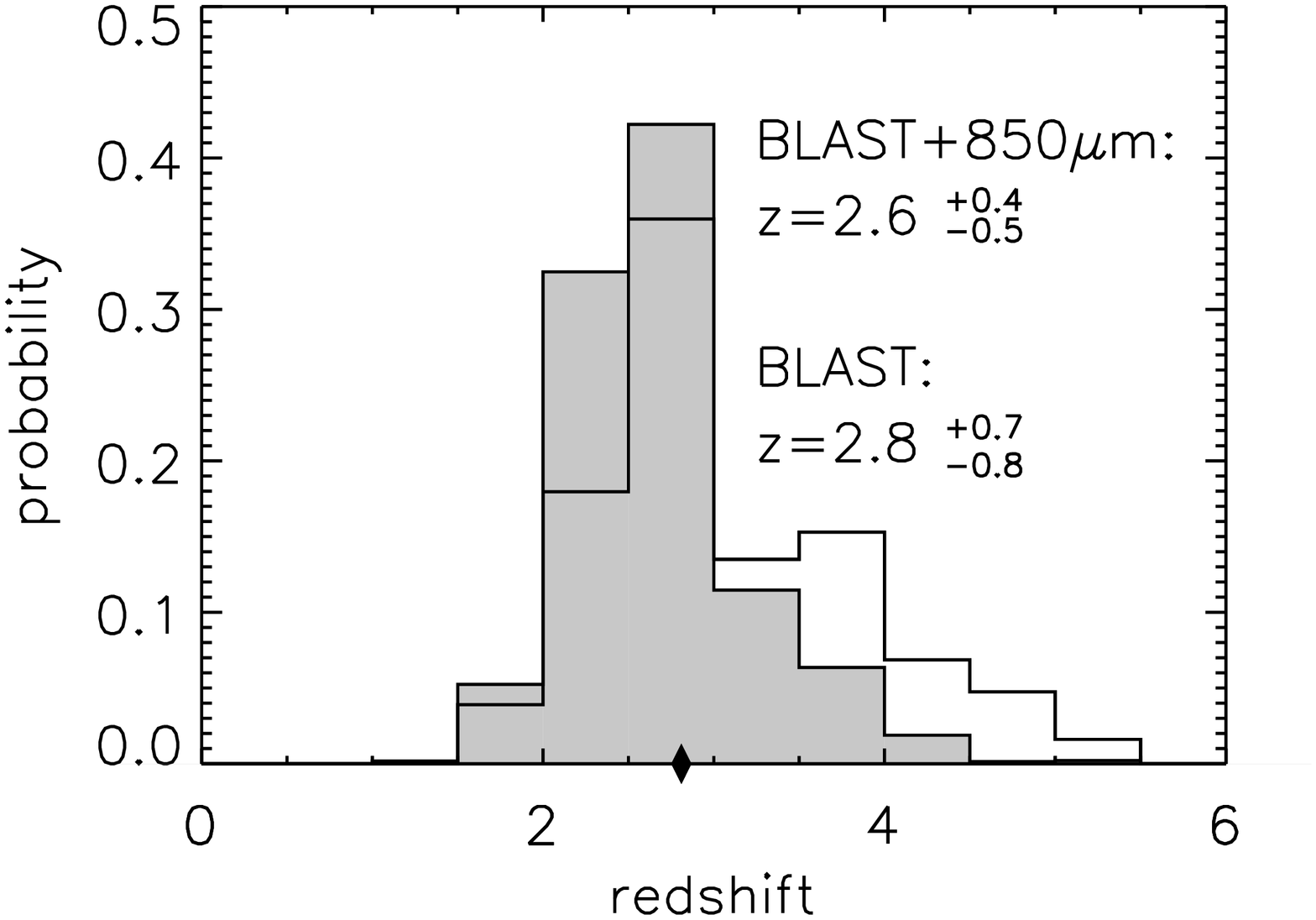}
  \includegraphics[height=0.4\textheight,angle=90]{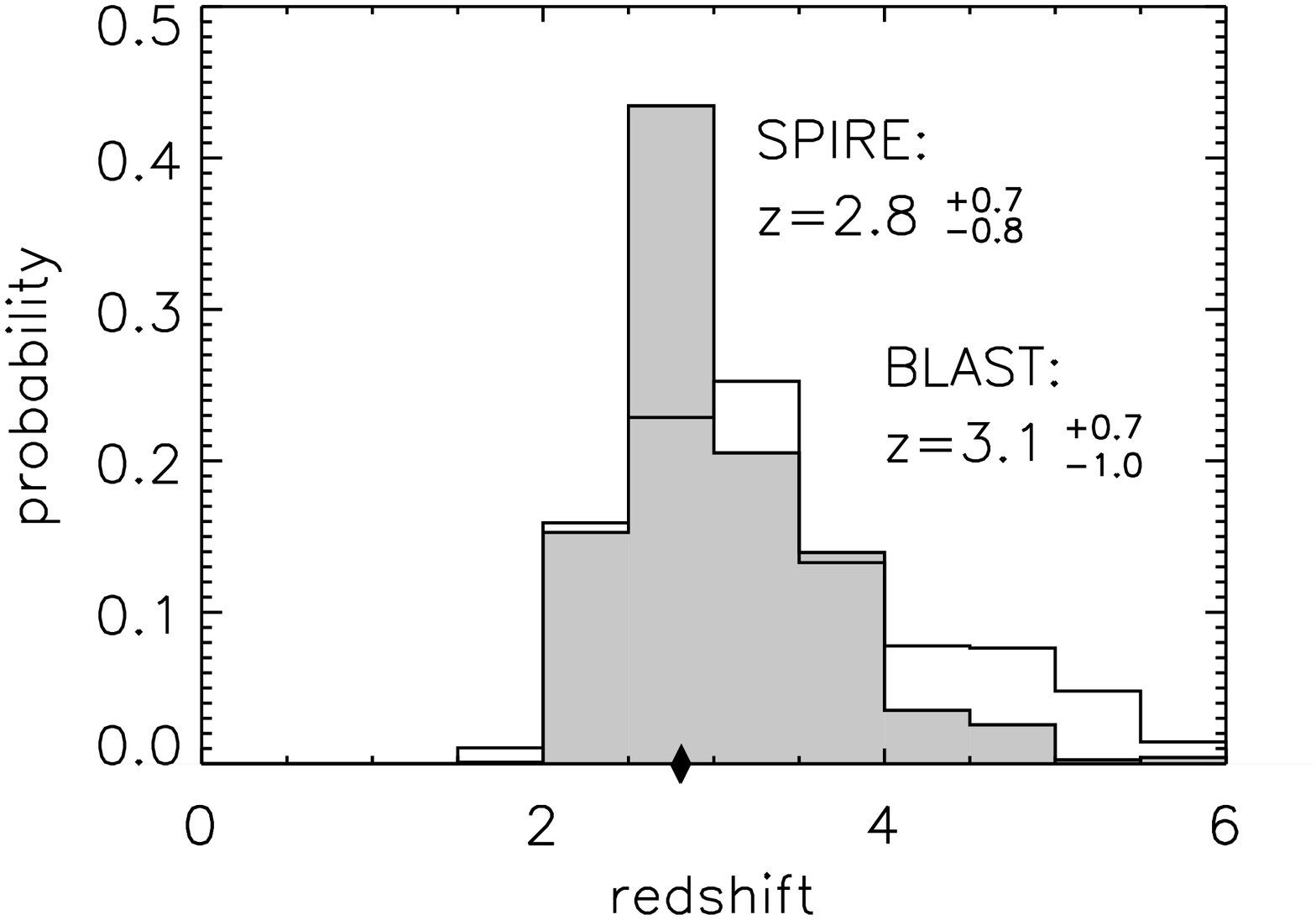}
  \caption{
{\it Left panel:} 
Redshift probability distributions of a $4\times 10^{12}$~\Lsun\ galaxy 
at $z=2.8$, detected at 250, 350 and 500$\mu$m with 
BLAST, and also at 850$\mu$m in a ground-based deep survey. The intrinsic 
colours of this galaxy are determined from
a scaled SED of NGC~1614: 
$S_{250\mu{\rm m}}=17$~mJy, $S_{350\mu{\rm m}}=23$~mJy,
$S_{500\mu{\rm m}}=27$~mJy, $S_{850\mu{\rm m}}=5$~mJy.
The unshaded  and shaded distributions have been calculated from 
BLAST detections and from BLAST+850$\mu$m detections, respectively. 
The black diamond marks the true redshift of the mock galaxy.
Photometric redshift determinations within a 68\% confidence level
are indicated within the panel.
{\it Right panel:} A comparison of the redshift distributions for a
$3\times 10^{12}$~\Lsun\ galaxy at $z=2.8$ determined from SPIRE and the
less sensitive BLAST observations.  
The colours of this galaxy are similar to NGC~2992:
$S_{250\mu{\rm m}}=18$~mJy, $S_{350\mu{\rm m}}=28$~mJy,
$S_{500\mu{\rm m}}=23$~mJy, $S_{850\mu{\rm m}}=7$~mJy.
The galaxy was detected simultaneously in the 3 SPIRE passbands, 
but only at 350 and 500$\mu$m with BLAST.
The unshaded  and shaded histograms correspond to the 
use of BLAST and SPIRE detections, respectively. 
}
\end{figure}
\begin{figure}
\hspace*{-0.5cm}
  \includegraphics[height=0.48\textheight,angle=90]{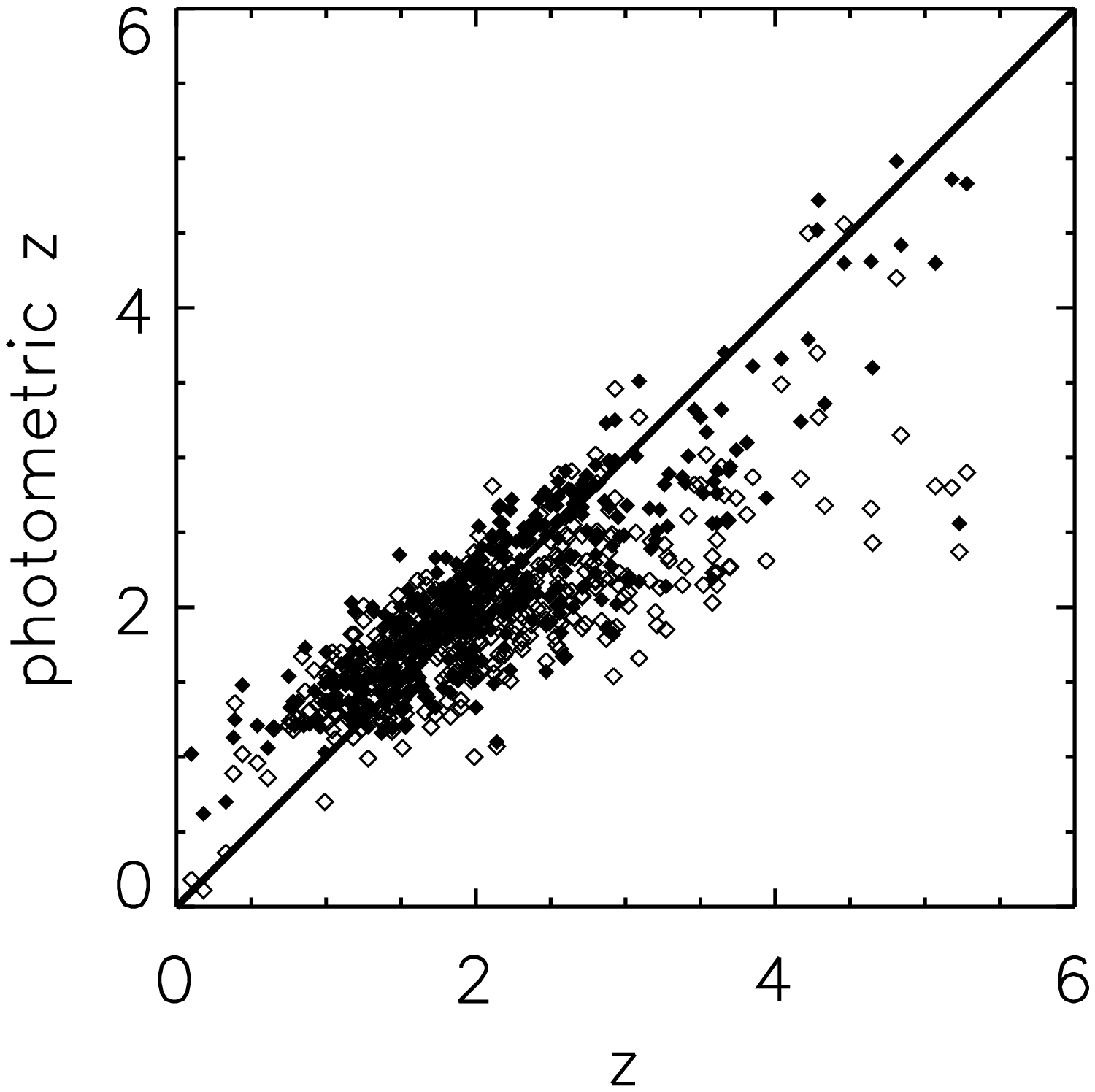}
\hspace*{-2cm}
  \includegraphics[height=0.4\textheight,angle=90]{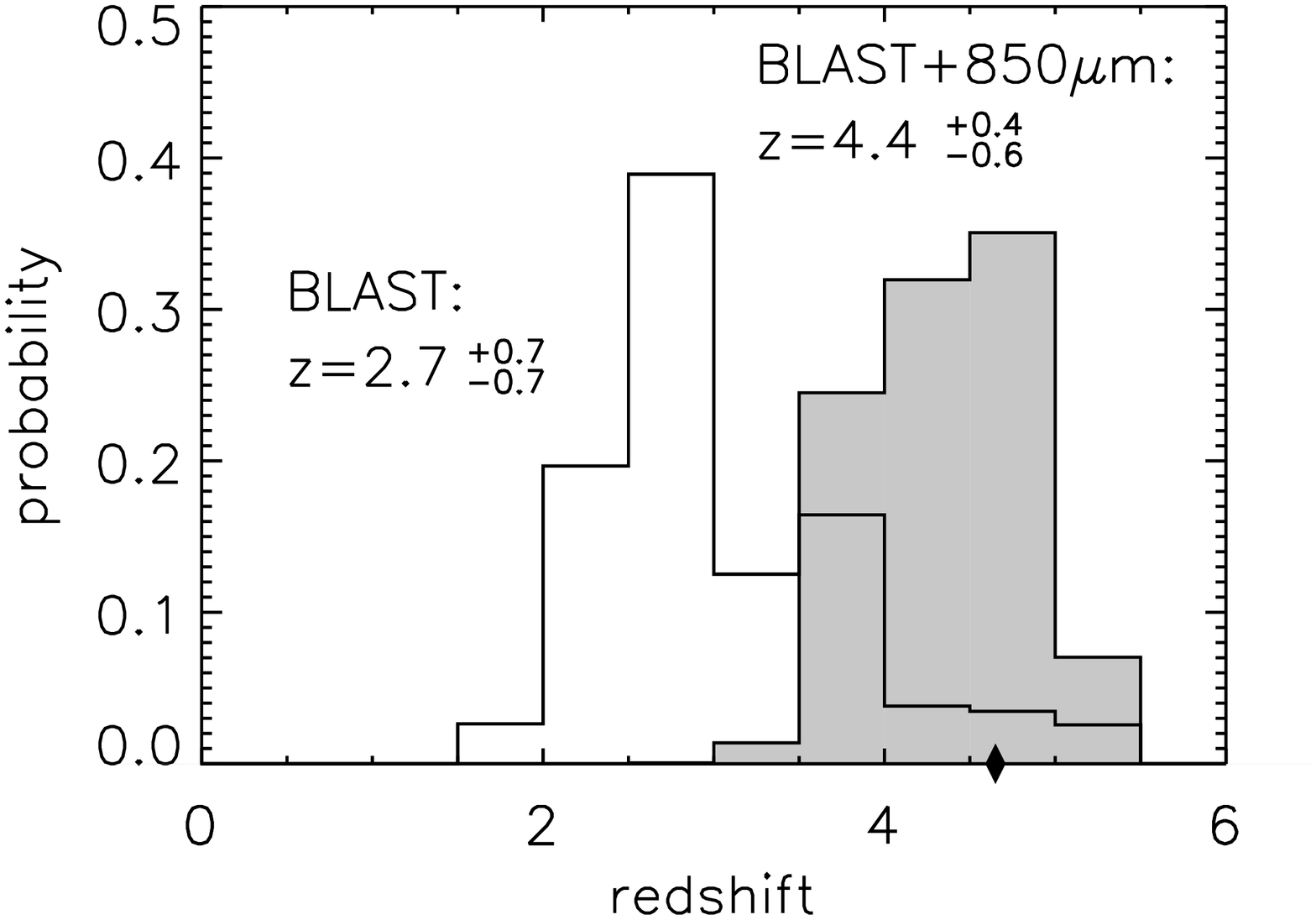}
  \caption{
{\it Left panel:} Photometric redshift vs. true redshift relationship
for 400 galaxies simultaneously detected at 250, 
350 and 500$\mu$m in 1 sq. degree.
Open symbols show the relationship inferred using only BLAST data to 
derive the redshifts. Filled symbols show the relationship when 
the redshifts are estimated using colours based on BLAST and complementary 
850$\mu$m detections from a ground-based survey.
 The addition of 850$\mu$m measurements significantly 
increases the accuracy at $z\gsim4$, since at these redshifts 
the BLAST filters sample the rest-frame mid-IR to FIR ($\sim$35--100$\mu$m).
The longer wavelength data are required to bracket the 
rest-frame FIR peak,
which provides the diagnostic power for the photometric redshift 
technique discussed in this paper. {\it Right panel:} Example of the 
correction attained at $z>4$ when 850$\mu$m observations are included in
the photometric redshift analysis. In particular, these are redshift 
distributions
of a $1\times 10^{13}$~\Lsun\ galaxy at  $z=4.65$, with observed fluxes
$S_{250\mu{\rm m}}=18$~mJy, $S_{350\mu{\rm m}}=29$~mJy,
$S_{500\mu{\rm m}}=27$~mJy, $S_{850\mu{\rm m}}=19.5$~mJy.
}
\end{figure}

\section{OPPORTUNITIES FOR SPECTROSCOPIC FOLLOW-UP}

Redshift accuracies can be further improved with heterodyne follow-up
observations. We plan to use the recently commissioned 100m
GBT to search for low-J CO molecular-line transitions. 
For example, for those BLAST and Herschel/SPIRE galaxies 
with photometric redshifts in the range $z \sim 4.0 - 4.8$, which have
1$\sigma$ accuracy of $ \Delta z \sim \pm 0.4$ when combined with 850$\mu$m 
observations, we can  search for CO(1--0) in the K-band (18.0 -- 26.5~GHz). 
The addition of a future GBT receiver operating at 26 -- 40~GHz will extend 
the CO(1--0) search 
to galaxies at $2.3 \lsim z \lsim 4.8$.  Similarly, a proposed 68 -- 116~GHz receiver, 
will detect both the CO(1--0) and CO(2--1) transitions of the low-$z$ 
population ($z \lsim 2$).

To conclude, our simulations demonstrate that,
within the next few years, large-area BLAST
surveys will produce a
catalogue of $\gsim 5000$ high-$z$ galaxies for which sub-mm photometric
redshifts with an accuracy $\Delta z \sim \pm 0.4-0.6$ can be
determined. Follow-up observations with GBT receivers will provide definitive 
molecular-line redshifts and dynamical-mass estimates.
The combination of BLAST and GBT provide a powerful combination to
break the redshift {\em deadlock} that hinders our ability to
understand the evolution and nature of the sub-mm starburst galaxy
population.

\vspace*{0.3cm}
\noindent
{\bf Acknowledgments:}
This work has been partly supported by CONACyT grants 32143-E
and 32180-E. 





\end{document}